\documentclass[aps,prd,twocolumn,superscriptaddress,showpacs,showkeys,nofootinbib]{revtex4}
\usepackage[dvips]{graphicx}
\usepackage[english]{babel}
\usepackage{xcolor}
\usepackage{mathtools}
\usepackage{pdfsync}

\usepackage{psfrag}

\usepackage{amssymb,epsf,epsfig}
\usepackage{amsmath}

\usepackage{graphicx}

\usepackage{slashed}
\usepackage[active]{srcltx}

\def \be  {\begin{equation}}
\def \ee  {\end{equation}}
\def \ba  {\begin{eqnarray}}
\def \ea  {\end{eqnarray}}
\def \baa {\begin{eqnarray*}}
\def \eaa {\end{eqnarray*}}
\def \lab #1 {\label{#1}}

\def\d{\hbox{{d}\kern-.20em\hbox{l}}}
\def \qqquad {\qquad\quad}
\def \qqqquad {\qquad\qquad}
\def \matrix #1 {\left(\begin{array}{cc} #1 \end{array}\right)}

\def \tr {\mathop{\rm tr}\nolimits}

\def\II{\hbox{{1}\kern-.25em\hbox{l}}}

\newcommand \widebar [1] {\overline{#1}}

\newcommand \vev [1] {\langle{#1}\rangle}

\begin{document}

\title{Unitarity violation in non-integer dimensional Gross-Neveu-Yukawa model}

\date{\today}

   \author{Yao Ji}
   \email{yao.ji@ur.de}
\affiliation{Institut f\"ur Theoretische Physik, Universit\"at
   Regensburg, D-93040 Regensburg, Germany}

   \author{Michael Kelly}
   \email{Michael.Kelly@stud.uni-regensburg.de}
\affiliation{Institut f\"ur Theoretische Physik,  Universit\"at
   Regensburg, D-93040 Regensburg, Germany}

   \date{\today}

\begin{abstract}
We construct an explicit example of unitarity violation in fermionic quantum field theories in non-integer dimensions. We
study the two-point correlation function of four-fermion operators. We compute the one-loop anomalous dimensions of these
operators in the Gross-Neveu-Yukawa Model. We find that at one loop order, the four-fermion operators split into three classes
with one class having negative norms. This implies that the theory violates unitarity following the definition in Ref.~\cite{Hogervorst:2015akt}.
\end{abstract}



\maketitle

\section{Introduction}

The conformal field theories (CFTs) have always been an area of active research
due to their rich mathematical structure and  physical applications.
In unitary theories conformal symmetry imposes severe constraints on the spectrum of operator dimensions.
It is believed  that these dimensions can be determined with the help of the conformal bootstrap technique~\cite{Polyakov:1974gs,
Ferrara:1973yt}. This technique proved to be extremely useful for solving  two dimensional CFTs. The effective
numerical algorithms for solving the
bootstrap equations for higher-dimensional CFTs have been proposed in Ref.~\cite{Rattazzi:2008pe},
(see also Refs.~\cite{Rychkov:2011et, ElShowk:2012ht, Kos:2013tga}, ~\cite{Rychkov:2009ij, Caracciolo:2009bx, Poland:2010wg,
Rattazzi:2010gj, Rattazzi:2010yc,
Vichi:2011ux}, and~\cite{Nakayama:2014yia, Li:2016wdp, Bae:2014hia, Poland:2011ey, ElShowk:2012hu}
for more details and recent developments in $d=3$, $d=4$ and $d=5$ dimensions, respectively). One of the advantages of this approach is that it allows one to obtain operator
dimensions directly in various integer dimensions.

The standard technique for the calculation of the operator dimensions, the so-called $\epsilon$-expansion~\cite{Wilson:1971dc,Wilson:1972cf},
is based on  calculation of the scaling dimensions in  $d=4-2\epsilon$ dimensional theory and interpolation of  the relevant critical
indices   to
the physical dimension. The critical indices for many CFTs are known with high precision. One of the recent
 achievements is the calculation
of the six-loop $\beta$ function in the $\varphi^4$ theory~\cite{Kompaniets:2017yct}. In order to get a better understanding of the new conformal
bootstrap technique it was quite natural to apply it to theories in non-integer dimensions, $d=4-2\epsilon$,
	see Ref.~\cite{El-Showk:2013nia, Liendo:2012hy}.
At the same point one of the assumptions which most of the conformal bootstrap relies on is the unitarity of the theory. One can hardly
expect that this assumption -- unitarity -- will be true for theories in non-integer dimensions. This question was raised in
Refs.~\cite{Hogervorst:2015akt, Hogervorst:2014rta} where  unitarity violation in $\varphi^4$ theory  was demonstrated by constructing states
(operators) with negative norm. The first ``negative norm'' operator in $\varphi^4$ theory  has a rather high scaling dimension
($\Delta=23$) and it is expected that unitarity breaking effects will appear only in high orders of $\epsilon$
expansion. Negative norm operators  have necessarily to be evanescent operators, i.e.  operators that are vanishing in integer dimensions.
In scalar theories the building blocks for the operators  are fields and their derivatives and therefore evanescent
operators
are obliged to have a high dimension. The situation is quite different in  theories with fermions where there are
evanescent (scalar) operators of canonical dimension six~\cite{Dugan:1990df}.

The aim of this article is to demonstrate the existence of the negative norm-states in the $d=4-2\epsilon$ dimensional
Gross~-~Neveu~-~Yukawa (GNY) model~\cite{ZinnJustin:1991yn}. It was argued in~\cite{Hogervorst:2015akt} that unitarity  implies  the
positiveness of the coefficient $C$ in the correlator
\begin{align}\label{eq:CFT}
\langle \mathcal{O}^\dagger(x)\mathcal{O}(0)\rangle=C/x^{2\Delta}\,,
\end{align}
where $\mathcal{O}$ is a conformal operator with scaling dimension $\Delta$. In an integer dimensional CFT,
violation of this condition indicates the presence of negative norm states in the theory~\cite{Hogervorst:2015akt}. We consider
the renormalization of an infinite set of scalar four-fermion operators in $d=4-2\epsilon$ dimensions and show that the positiveness condition is broken
for infinitely many operators.
Since the canonical dimension of these operators is not large,
$\Delta^{\text{can}}=6$, one can wonder about the effect of negative norm operators to the conformal bootstrap technique.

	The article is organized as follows:
	In section II we discuss the two-point correlation function of scalar four-fermion operators in the free theory. We find that the theory contains evanescent operators which could generate negative norm states.

In order to continue our discussion we then compute in section III the anomalous dimension of the physical and evanescent operators at one loop-order in the GNY model. It turns out that all the evanescent operators split into two classes of definite anomalous dimension. We show that the negative norm states are generated by one of this two classes, depending on the number of fermion flavors of the theory.

\section{Four-fermion correlation function in non-integer dimensions}

The GNY model describes an interacting fermion-boson system with the Lagrangian given by the following
expression~\cite{ZinnJustin:1991yn, Hasenfratz:1991it}
\begin{align}
{\cal L}&=\frac12\left(\partial_\mu\sigma\right)^2+\bar\Psi_i\slashed{\partial}\Psi_i+g_1\sigma\bar\Psi_i\Psi_i+\frac1{24}g_2\sigma^4\,,
\end{align}
where the index $i=1,\ldots,n_f$ enumerates different fermion flavors and $\sigma$ is a scalar field.

The model has an infrared stable fixed point in $d=4-2\epsilon$ dimensions~\cite{Moshe:2003xn}.  At one loop the critical couplings take the form
\begin{align}\label{crit-p}
u_*&=\frac{(g_1^*)^2}{(4\pi)^2}=\frac{\epsilon}{N_f+6}\, ,\notag\\
v_*&=\frac{(g_2^*)^2}{(4\pi)^2}=\frac{6-N_f+\sqrt{N_f^2+132N_f+36}}{6(N_f+6)} \, \epsilon,
\end{align}
where  $N_f\equiv n_f\,\tr({\mathbb I}_d)$. The basic critical indices are known now with four loop accuracy and can be found
in Ref.~\cite{Zerf:2017zqi}.

Let us consider an infinite system of four-fermion local operators in $d=4-2\epsilon$ dimensions
\begin{align}
\label{eq:O}
{\cal O}^{(m)}
&=\frac1{m!}\Big(\bar\Psi\,\Gamma^{(m)}_{\mu}\Psi\Big)\Big(\bar\Psi\,\Gamma_{(m)}^\mu\Psi\Big)\, .
\end{align}
A summation over flavor index inside each bracket is tacitly assumed.  
The notation $\Gamma^{(m)}_{\mu}$ stands for an antisymmetric product of $m$ $\gamma$-matrices
\begin{align}
\label{eq:Gammadefs}
\Gamma^{(m)}_\mu&=\Gamma_{\mu_1\ldots\mu_m}\equiv
\frac1{m!}\sum_{s\in S_m} (-1)^P\gamma_{\mu_{s_1}}\ldots\gamma_{\mu_{s_m}}\, .
\end{align}
The sum goes over all permutations 
and $P$ is the parity of a permutation. 

Before taking a closer look at correlators of the operators~\eqref{eq:O} let us state a few things about the $\Gamma_{(m)}$ matrices.
The Dirac $\gamma$-matrices satisfy
the basic anti-commutation relation in $d$-dimensional space
\begin{align}
\left\{\gamma^\mu,\gamma^\nu\right\}=2g^{\mu\nu}\,\mathbb{I}_d\, ,\qqqquad
g^{\mu\nu}g_{\mu\nu}=d\, ,
\end{align}
where $g^{\mu\nu}$ is the metric tensor. 
In integer dimensions there are only $d$ distinct gamma matrices $\gamma^0,\ldots,\gamma^{d-1}$.
This restricts the maximum number of different anti-symmetrized matrices~\footnote{Note that in even dimensions, $\Gamma_{(m>d)}$ vanishes because of the antistmmetrization of gamma matrices. In odd dimension $d$, $\Gamma_{(d)}$ is removed from the independent basis since $\Gamma_{(d)}\propto\Gamma_{(0)}$.} $\Gamma_{(m)}$. Namely, $0\leq m\leq d (\leq d-1)$ for even (odd) dimensional spaces.

In non-integer dimensions however, the situation  is different. There exists an infinite number of $\gamma$-matrices and therefore
it is possible to construct infinitely many non-vanishing and distinct $\Gamma^{(m)}$. As a result, the parameter $m$ in
Eq.~(\ref{eq:O}) takes any positive integer values. However, in $d=4-2\epsilon$ dimensional space the operators~\eqref{eq:O} with
$m\geq5$ have to vanish in the limit $\epsilon\to 0$ and therefore they are called evanescent operators.

The  renormalized operators $[\mathcal{O}_m]$ satisfy the renormalization group equation
\begin{align}
\left(M\partial_M + \beta_u \partial_u +\beta_v\partial_v \right)[\mathcal{O}_m]=-{\boldsymbol\gamma}_{\cal O}^{m,n}(u,v) [\mathcal{O}_n]\,,
\end{align}
where $M$ is the renormalization scale, $\beta_{u,v}$  are the corresponding $\beta$-functions,
$\beta_u=\frac{du}{d\ln M}$, $\beta_v=\frac{dv}{d\ln M}$ and
${\boldsymbol\gamma}_{\cal O}^{m,n} $ is the anomalous dimension matrix.
The structure of the operator mixing of the four-fermion operators was considered in great detail in \cite{Dugan:1990df,Herrlich:1994kh, Collins:1984xc}.

At the critical point  $\beta_u(u_*,v_*)=\beta_v(u_*,v_*)=0$ the problem of constructing operators with autonomous scale dependence
is equivalent to the eigenproblem for the matrix ${\boldsymbol\gamma}_{\cal O}^{m,n} $. This means, if $c_\gamma^m$ is the left eigenvector
of the anomalous dimension matrix
\begin{align}
c_{\gamma}^m{\boldsymbol\gamma}_{\cal O}^{m,n}(u_*,v_*)=\gamma \,c_{\gamma}^n,
\end{align}
then the operator $\mathcal{O}_\gamma=\sum_m c_\gamma^m [\mathcal{O}_m]$ has an autonomous scale dependence
\begin{align}
\big(M\partial_M + \gamma\big) \mathcal{O}_\gamma=0\,.
\end{align}
The operator $\mathcal{O}_\gamma$ transforms in a proper way under conformal transformations and according to a general theory
the correlators of operators with different scaling dimensions ($\Delta_\gamma=6-4\epsilon+\gamma$) vanish, i.e.
\begin{align}\label{O-gamma}
\langle \mathcal{O}^\dagger_\gamma(x) \mathcal{O}_{\gamma'}(0)\rangle=\delta_{\gamma\gamma'}C_\gamma/x^{2\Delta_\gamma}\,.
\end{align}
In an unitary theory the coefficients $C_\gamma$ have to be positive~\cite{Hogervorst:2015akt}.
We calculate the one-loop anomalous dimension matrix $\boldsymbol\gamma_{\cal O}^{m,n}$ in the next section while in the rest of this section we study the correlator~\eqref{O-gamma} in more detail.

Let us write the correlator~\eqref{O-gamma} in the form
\begin{align}
\langle \mathcal{O}^\dagger_\gamma(x) \mathcal{O}_{\gamma'}(0)\rangle=\sum_{m,n} (c_\gamma^m)^\dagger
C^{m,n}(x) c^n_{\gamma'} ,
\end{align}
where $C^{m,n}$ is the correlator of the basic operators defined in Eq.~\eqref{eq:O} (note that $(\mathcal{O}_n)^\dagger=\mathcal{O}_n$ and $d=4-2\epsilon$)
\begin{align}
\label{eq:correlator}
C^{m,n}(x) &=\vev{{\cal O}^{(m)}(x){\cal O}^{(n)}(0)}\notag\\
&= \frac{{\cal C}^{m,n}(d) }{|x^2|^{2d-2}}\Big(1+{\cal O}(u_*,v_*)\Big) \, 
\end{align}
In $d=4-2\epsilon$ dimensions, it is expected that for the physical operators ($m,n\leq 4$), ${\cal C}^{m,n}(d)\sim {\cal O}(1)$ and if one of the indices $m,n\geq5$, ${\cal C}^{m,n}(d)\sim
{\cal O}(\epsilon)$. Thus one gets the following expression for the constant $C_\gamma$ at the leading order
\begin{align}
C_\gamma=\sum_{n,m} (c_\gamma^m)^\dagger
C^{m,n}(x)c^n_{\gamma} \equiv (c_\gamma, Cc_\gamma)\,.
\end{align}

\begin{figure}
         \psfrag{t1}[cc][bc]{$T_1^{m,n}$}
         \psfrag{t2}[cc][bc]{$T_2^{m,n}$}
         \psfrag{x}[cc][cc]{\small$x$}
         \psfrag{0}[cc][cc]{\small$0$}
\includegraphics[width=1.0\linewidth]{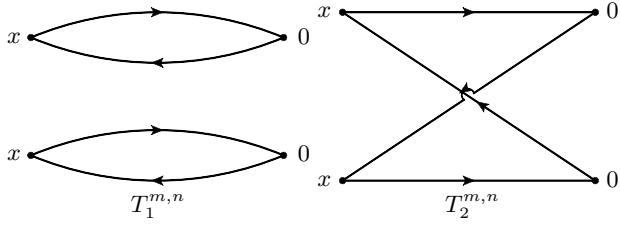}\\\vspace{.2cm}
 \caption{\label{fig:loops}Feynman diagrams of leading order}
\end{figure}

At leading order only the two Feynman diagrams shown in Fig.~\ref{fig:loops} contribute to $C^{m,n}(d)$. %
%
%
Using the expression for the fermion propagator in Euclidean space
 \begin{align}
 \label{eq:propagator}
 \vev{\Psi(x)\bar\Psi(0)}= A \frac{\slashed{x}}{{|x^2|}^{d/2}}\, , && A=\frac{\Gamma(d/2)}{2\pi^{d/2}}\,
 \end{align}
  we find
 \begin{align}
 \label{eq:CT1T2}
 C^{m,n}(x)&= \Delta^{m,n}{ A^4 N_f}{{|x^2|}^{2-2d}} \, \, ,
 \end{align}
 where
 \begin{align}
 \Delta^{m,n}&=N_f \,T_1^{m,n}+T_2^{m,n}\, ,\notag\\
  T_1^{m,n}&=\frac{\delta_{m,n}}{x^4\tr^2(\mathbb{I}_d)(m!)^2}\left[\tr\big(\Gamma^{(m)}_\mu\slashed{x}\,\Gamma^{(m)}_\nu\slashed{x}\big)\right]^2,\notag\\
   T_2^{m,n}&=\frac{-1}{x^4\tr(\mathbb{I}_d)m!n!}\tr\big(\slashed{x}\,\Gamma^{(m)}_\mu\slashed{x}\,\Gamma^{(n)}_\nu\slashed{x}\,\Gamma_{(m)}^\mu\slashed{x}\,\Gamma_{(n)}^\nu\big)\, .
\label{eq:DetlaT}
\end{align}
The summation between upper and lower indices is here implied.
The calculation of the traces in~\eqref{eq:DetlaT} is discussed in the Appendix (see Ref.~\cite{Vasiliev:1995qj} for a general treatment of contracting infinitely many anti-symmetrized gamma matrices), here we present the final result~\footnote{A nontrivial relation in integer dimensions $\tr^2(\mathbb{I})[T_1T_2^{-1}]^2={\mathbb I}$ serves as an additional check for our results of $T_1^{m,n}$ and $T_2^{m,n}$. This relation is obtained from considering the Fierz identities.}
 \begin{align}
\label{eq:T1final}
T_1^{m,n}&=\frac{\Gamma(d+1)}{m!\,\Gamma(d-m+1)}\,\delta^{m,n}\, ,\\
\label{eq:T2final}
T_2^{m,n}&=-\frac12\,i^{m(m+1)+n(n+1)}\,a^{m,n}\, .
 \end{align}
Note that $T_1^{m,n}$ and $T_2^{m,n}$ are $x$-independent.
The coefficients $a^{m,n}$ are encoded by the generating function
\begin{align}
\label{eq:FExpand}
F(x,y)&=\sum_{m,n=0}^\infty a^{m,n}\,x^my^n
\notag\\
&= (1-x+y+xy)^d+(1+x-y+xy)^d
\notag\\&\quad-(1+x+y-xy)^d
+(1-x-y-xy)^d\, .
\end{align}


\begin{figure}
         \psfrag{d}[cc][cc]{$d$}
          \psfrag{Nt}[cc][cc]{\rotatebox{180}{$\Delta^{m,m}(d)$}}
 \includegraphics[width=0.8\linewidth]{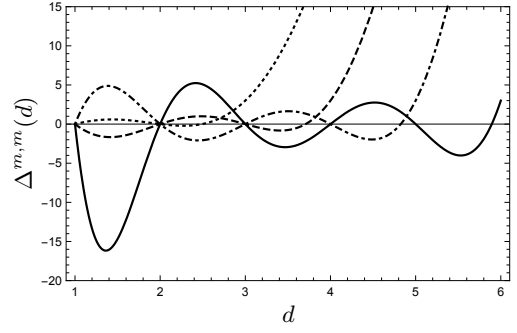}\\\vspace{.2cm}
 \caption{\label{fig:C} $\Delta^{m,m}(d)$ with $N_f=4$ as function of space dimension $d$ for $m=3,4,5,6$ presented by dotted,
 dashed, dot-dashed and  solid curves, respectively.}
\end{figure}

We point out that $T_1^{m,n}$ and $T_2^{m,n}$ are symmetric regarding the exchange of $m\leftrightarrow n$ and in contrast to the first diagram, which is proportional to $\delta^{m,n}$, $a^{m,n}$
contributes to both cases of $m=n$ and $m \neq n$. Both diagrams are
polynomials in the spacetime dimension $d$ and can become
negative valued in non-integer dimensions. Therefore the coefficient $\Delta^{m,n}$ is negative valued in some regions (see Fig. \ref{fig:C}).  A detailed analysis of $a^{m,n}$ shows that $|T^{m,m}_2| \gg |T^{m,m}_1|$
for $m \gg 1$ and therefore gives the main contribution at large $m$ for $\Delta^{m,m}$. 
The fact that $\Delta^{m,m}$ ($\sim C^{m,m}$) can become negative valued suggests the possibility of having conformal operators with negative norms. For this reason we compute the one-loop anomalous dimension of the operators ${\cal O}^{(m)}$ in the next section in order to classify them by their one-loop anomalous dimensions.

\section{Anomalous dimension and unitarity in the GNY model}

\begin{figure}
\begin{minipage}{0.43\textwidth}
         \psfrag{d}[cc][cc]{$d$}
          \psfrag{Nt}[cc][cc]{\rotatebox{180}{$\Delta^{m,m}(d)$}}
 \includegraphics[width=0.8\linewidth]{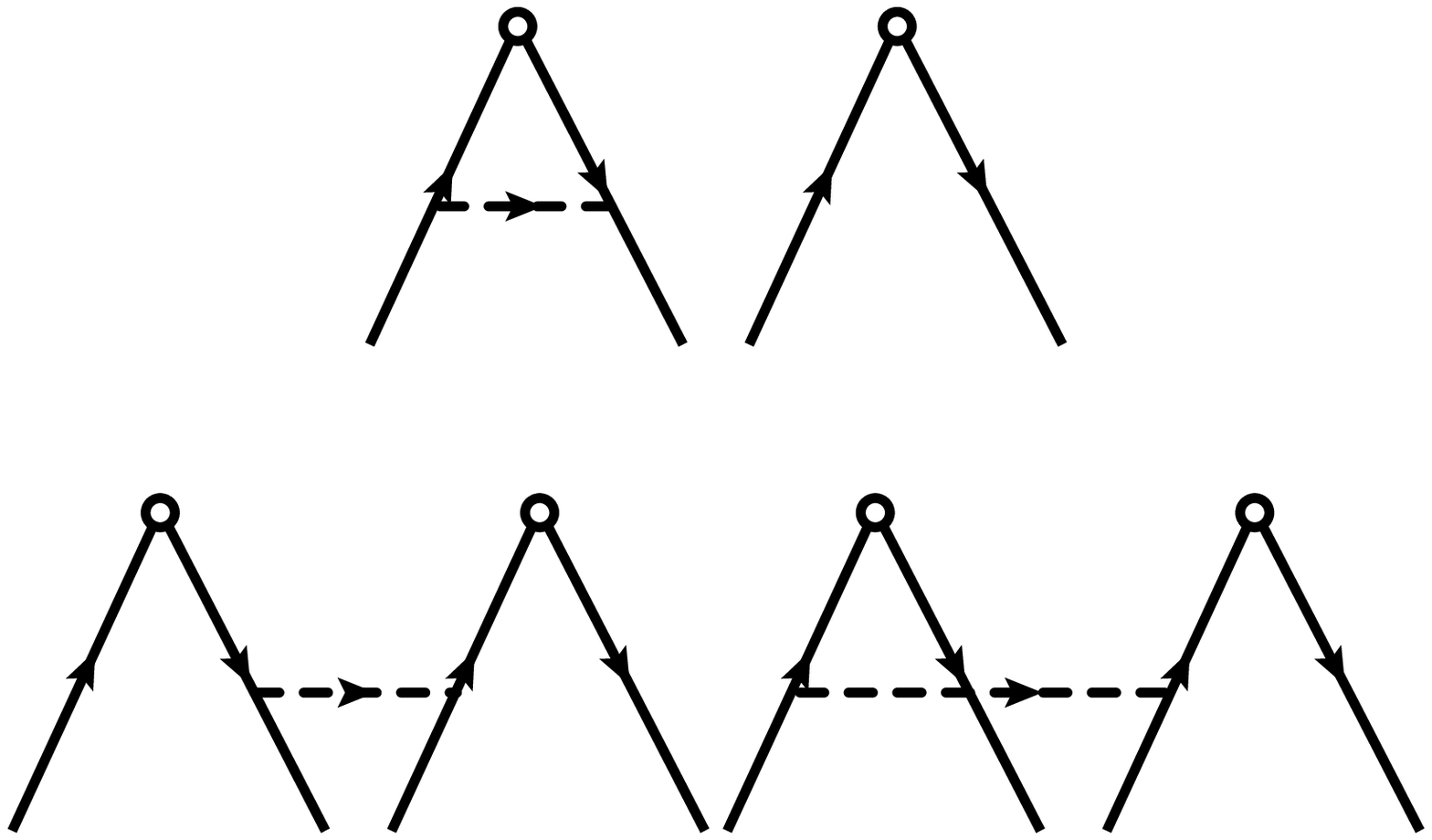}\\\vspace{.2cm}
 \caption{\label{fig:AMD} Feynman diagrams for calculating anomalous dimensions of evanescent operators.}
\end{minipage}
\vspace{10pt}


\begin{minipage}{0.43\textwidth}
 \includegraphics[width=1.0\linewidth]{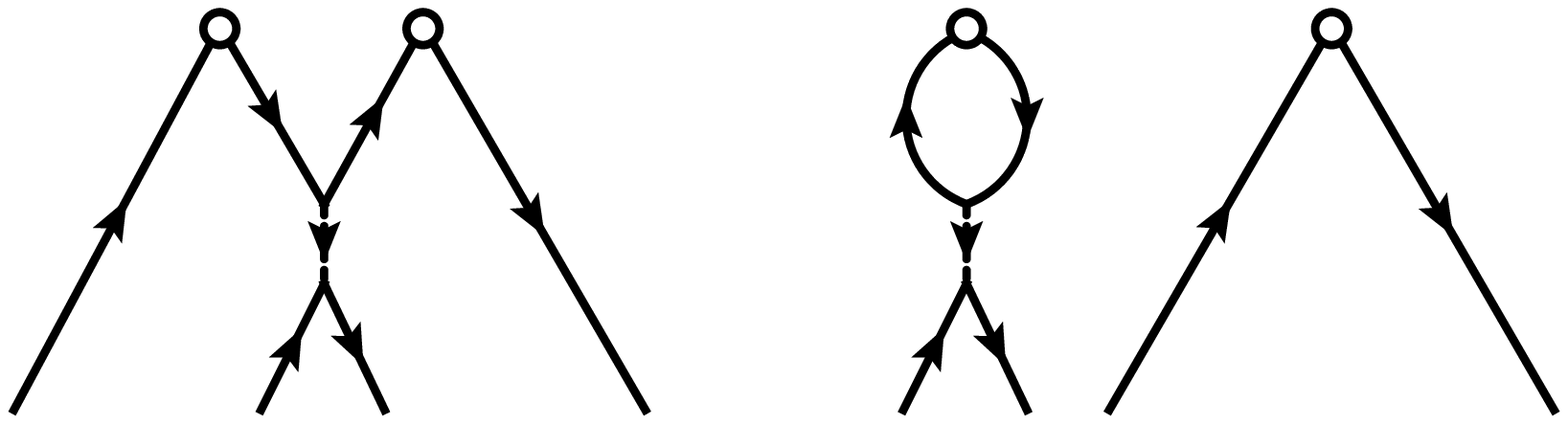}\\\vspace{.2cm}
 \caption{\label{fig:AMDphy} Additional Feynman diagrams contributing to the anomalous dimensions of physical operators.}
\end{minipage}
\end{figure}

So far our calculations are rather general and can be applied to any fermionic theory in non-integer dimensions. In order to
proceed our study of norm states in a conformal theory, according to Eq.~\eqref{O-gamma}, 
it is necessary to find eigenstates with definite anomalous dimensions
and study correlation functions between them. It is therefore more instructive to consider an explicit example, the GNY
model, and compute the one-loop anomalous dimensions of the operators ${\cal O}^{(m)}$ defined in Eq.~\eqref{eq:O} in this model.
The Feynman diagrams needed for this calculation are given in Figs.~\ref{fig:AMD} and~\ref{fig:AMDphy}. Note that diagrams in
Fig.~\ref{fig:AMDphy} only contribute to the anomalous dimension of physical operators.

Then it is straightforward to compute these one-loop diagrams and obtain the anomalous dimension matrix
${\boldsymbol\gamma}_{\cal O}^{m,n}$. Interestingly, we find that the anomalous dimension matrix has a simple block diagonal form
(the calculation details can be found in Appendix~\ref{sec:AD}),
\begin{align}
{\boldsymbol\gamma}_{\cal O}^{m,n}=2u_*\,
{\rm diag}({\boldsymbol\gamma}_0\, , {\boldsymbol\gamma}_1\, , {\boldsymbol\gamma}_2\, ,\ldots)^{m,n}\, ,
\end{align}
where $\boldsymbol\gamma_0$ is a $5\times5$ anomalous dimension matrix
\begin{align}
\boldsymbol\gamma_0=
\begin{pmatrix}
N_f+2 & 0 & 0 & 0 & 0\\
-4 & 0 & -2 & 0 & 0\\
6 & -3 & 1 & 0 & 0\\
4 & 0 & 0 & 2 & -4\\
-1 & 0 & 0 & -1 & -1
\end{pmatrix}
\end{align}
involving only physical operators, while $\boldsymbol\gamma_{k\geq1}$ are $2\times2$ matrices
\begin{align}
\boldsymbol\gamma_k&=
\begin{pmatrix}
2k+2 & -2k-4\\
2k-1 & -2k-1
\end{pmatrix}
\end{align}
describing the mixing between evanescent operators ${\cal O}^{(2k+3)}$ and ${\cal O}^{(2k+4)}$ at the one-loop order.

It is clear from the explicit expression of the anomalous dimension matrix that the physical and evanescent operators decouple
at the one-loop order. We can therefore study them separately and find the conformal basis in each case.

Let us write the physical operators in conformal basis as $\widetilde{\cal O}$. Then 
\begin{align}
\begin{pmatrix}
\widetilde{\cal O}_{0}^{(0)}\vspace{0.8mm} \\
\widetilde{\cal O}_+^{(1)} \vspace{0.8mm}\\
\widetilde{\cal O}_{-}^{(1) \vspace{0.8mm}}\\
 \widetilde{\cal O}_+^{(2)} \vspace{0.8mm}\\
 \widetilde{\cal O}_{-}^{(2)}
 \end{pmatrix}=
 \begin{pmatrix}
 1 & 0 & 0 & 0 & 0 \vspace{0.8mm}\\
 \frac{10}{1-N_f} & -1 & 1 & 0 & 0 \vspace{0.8mm}\\
 0 & 3/2 & 1 & 0 & 0 \vspace{0.8mm}\\
 \frac{5}{N_f-1} & 0 & 0 & -1 & 1 \vspace{0.8mm}\\
 0 & 0 & 0 & 1/4 & 1
 \end{pmatrix}
\begin{pmatrix}
{\cal O}^{(0)} \vspace{0.8mm}\\
{\cal O}^{(1)} \vspace{0.8mm}\\
{\cal O}^{(2)} \vspace{0.8mm}\\
{\cal O}^{(3)} \vspace{0.8mm}\\
{\cal O}^{(4)} \vspace{0.8mm}\\
\end{pmatrix}
\end{align}
where the operators $\widetilde{\cal O}^{(0)}_0$, $\widetilde{\cal O}^{(k)}_+$, and $\widetilde{\cal O}^{(k)}_-$ have anomalous
dimension $\gamma_0=2(N_f+2) u_*$, $\gamma_+=6u_*$, and $\gamma_-=-4u_*$ at one loop order, respectively. Note that we use the bold font letters for anomalous dimension matrices and common ones for the eigenvalues.

The conformal basis for evanescent operators, denoted as $\widebar{\cal O}^{(k)}_\pm$, is
\begin{align}
\begin{pmatrix}
\widebar{\cal O}^{(k)}_+ \vspace{0.8mm}\\
\widebar{\cal O}^{(k)}_-
\end{pmatrix}=
\begin{pmatrix}
-1 & 1\vspace{0.8mm} \\
\frac{1-2k}{2(k+2)} & 1
\end{pmatrix}
\begin{pmatrix}
{\cal O}^{(2k+3)} \vspace{0.8mm}\\
{\cal O}^{(2k+4)}
\end{pmatrix}
\label{eq:evanbasis}
\end{align}
with $k\geq1$ and $\widebar{\cal O}^{(k)}_\pm$ having anomalous dimension $6u_*$ and $-4u_*$,
respectively.
This results allow us to classify the operators by their one-loop anomalous dimension. More explicitly, the evanescent operators form two and the physical operators form three (two for $N_f=1$) classes. At this point one should mention that the two-loop anomalous dimensions of the operators ${\cal O} ^{(m)}$ probably allow us to do further classifications. The anomalous dimensions of the different operators are collected in the table~\ref{table:adconformaloperators}.

\begin{table}[h]
  \begin{center}

  \begin{tabular}{ l | c | c | c | c | c }

     &   $\widetilde{\cal O}^{(0)}_0$
      & $\widetilde{\cal O}^{(k)}_+$ & $\widetilde{\cal O}^{(k)}_-$ & $\widebar{\cal O}^{(k)}_+$ & $\widebar{\cal O}^{(k)}_-$\\ \hline
     \rule{0pt}{3ex}
    AD & $\gamma_0$ & $\gamma_+$ &  $\gamma_-$ & $\gamma_+$ & $\gamma_-$

    \end{tabular}

    \end{center}
    \caption{Anomalous dimension (AD) of the different physical and evanescent operators.}
    \label{table:adconformaloperators}
\end{table}

In order to find the negative norm states of the theory we have to consider
correlation functions between operators of the same anomalous dimension.
According to Eq.~\eqref{O-gamma}, this corresponds to the study of
the coefficient $C_\gamma$. We point out that at one loop accuracy, the orthogonality condition in Eq.~\eqref{O-gamma}  is realized by the following expressions
\begin{align}
\vev{(\widetilde{\cal O}_{0,\pm}^{(k_1)}(x))^\dagger\widetilde{\cal O}^{(k_2)}_{\mp}(0)}&={\cal O}(\epsilon)\, ,\notag\\
\vev{(\widetilde{\cal O}_{0,\pm}^{(k_1)}(x))^\dagger\widebar{\cal O}^{(k_2)}_{\mp}(0)}&={\cal O}(\epsilon)\, , \notag\\
\vev{(\widebar{\cal O}_\pm^{(k_1)}(x))^\dagger\widebar{\cal O}^{(k_2)}_{\mp}(0)}&={\cal O}(\epsilon^2)\, ,
\end{align}
which is exactly what we find from Eqs.~\eqref{eq:CT1T2},~\eqref{eq:T1final} and~\eqref{eq:T2final}.

With the orthogonality condition checked at the one-loop order, let us now focus on the evanescent operators in the conformal basis. We write the correlator as
\begin{align}
\label{eq:ADcorrelator}
&\vev{(\widebar{\cal O}_\pm^{(k_1)}(x))^\dagger\widebar{\cal O}^{(k_2)}_\pm(0)}=A^4\frac{N_f}{|x^2|^{2d-2}}\widebar\Delta^{k_1,k_2}_\pm+{\cal O}(\epsilon^2)\, ,\notag\\
&\widebar\Delta_\pm^{k_1,k_2}=N_f T_{1\pm}^{k_1,k_2}+T_{2\pm}^{k_1,k_2}\, .
\end{align}
Here both $T_{1\pm}$ and $T_{2\pm}$ are proportional to $\epsilon$ and correspond to the first and the second diagram in Fig.~\ref{fig:loops}, respectively. $A$ is defined in Eq.~\eqref{eq:propagator}.

The matrices $T_{1\pm}$ are diagonal matrices.   It is easy to see that all matrix elements of  $T_{1-} (T_{1+})$ are positive
(negative) numbers. This implies that $(f, T_{1-}f )>0$ and $(f,T_{1+} f)<0$ for arbitrary nonzero vectors $f$, i.e.
$T_{1-}$ ($T_{1+}$) is a positive (negative) definite matrix. The situation with the matrices $T_{2\pm}$ is a bit more complicated since
they are not diagonal.
But we checked numerically~\footnote{It is possible to show that all the leading principal minors of $T_{2+}$ are positive while the $k$-th order leading minor of $T_{2-}$ is negative (positive) for odd (even) $k$.}
 that all truncated matrices $T_{2\pm}^N = (T_{2\pm})^{n,m}$ with $ n,m\leq N$ are positive definite
($T_{2+}$) and negative definite ($T_{2-}$) matrices for $N\leq 80$. The definiteness of $T_{1,2 \pm}$ implies

\begin{enumerate}
\item
In the large $N_f$ limit, the matrices $\widebar\Delta_{\pm}\sim T_{1\pm}(1+O(1/N_f))$ and therefore
\newline
\newline
$\widebar\Delta_+$ is negative definite
\newline
$\widebar\Delta_-$ is positive definite.
\item Contrary, for small values of $N_f$ ($N_f \lesssim 5)$, $|T_{2\pm}|$ dominates over
$|T_{1\pm}|$ and
\newline
\newline
$\widebar\Delta_+$ is positive definite
\newline
$\widebar\Delta_-$ is negative definite.
\end{enumerate}
As we have seen, the one loop corrections are not enough to resolve the operator mixing since infinitely many operators have the
same anomalous dimension at one loop. Nevertheless, it allows one to argue that a general conformal operator
with anomalous dimension $\gamma_+=6u_* + O(\epsilon^2)$ has the form~\footnote{
For the sake of clarity one should mention that the mixing with physical operators is here neglected. But it can be shown that the effect of the physical operators is at order ${\cal O}(\epsilon^2)$ by considering the orthogonality condition of the conformal operators. More specifically, one finds that the coefficients in front of the physical operators are at ${\cal O}(\epsilon)$ order.}

\begin{align}
\mathcal{O}_{\gamma_+}= \sum c^+_i \widebar{\cal O}_+^{(i)} + \sum c^-_j \widebar{\cal O}_-^{(j)}
\end{align}
where the coefficients $ c_i^+\sim O(1),$ while $c_k^-\sim O(\epsilon)$. One can easily see that the coefficient
$C_{\gamma_+},$ corresponding to the correlator of such operators, is given by
\begin{align}
C_{\gamma_+}= A^4\frac{N_f}{|x^2|^{2d-2}} (c^+, \widebar\Delta_+ c^+) + O(\epsilon^2)\, .
\end{align}
Similarly, for  a conformal operator with anomalous dimension $\gamma_- =-4u_* + O(\epsilon^2)$ one gets
\begin{align}
C_{\gamma_-}= A^4\frac{N_f}{|x^2|^{2d-2}} (c^-, \widebar\Delta_- c^-) + O(\epsilon^2)\, .
\end{align}
As we have shown, the coefficients $C_{\gamma_+}$ and $C_{\gamma_-}$ have opposite signs at order ${\cal O}(\epsilon)$~\footnote{This is because $\widebar\Delta_\pm$ are order ${\cal O}(\epsilon)$.} for either small $N_f$ or $N_f\to\infty$. Therefore, one class of the
operators inevitably generates the negative norm states of the theory, according to the criteria given in Ref.~\cite{Hogervorst:2015akt}. In particular, at the lower bound of $N_f=1$, all negative norm states are generated by $\widebar{\cal O}_-^{(i)}$. In this case, the fermion field has only one degree of freedom and the GNY model may become supersymmetric as suggested in Ref.~\cite{Fei:2016sgs}.

We then conclude that the negative norm states are an integral part of the GNY model in $d=4-2\epsilon$ dimensions. At one
loop-order, all negative norms states are generated by operators with anomalous dimension $\gamma_-$ for
$N_f\lesssim{\cal O}(1)$. We believe the negative norm states exist in other theories with fermionic degrees of freedom in
non-integer dimensions as well because Eqs.~\eqref{eq:CT1T2},~\eqref{eq:T1final} and~\eqref{eq:T2final} are valid for any
fermionic theory with any number of flavors.

\section{Conclusion}


We have demonstrated the existence of negative norm states in the Gross-Neveu-Yukawa model in $d=4-2\epsilon$ dimensions
through the study of the two-point correlation functions of four-fermion operators and their one-loop anomalous dimension
matrix. The negative norm states we found are unavoidable as the two-point correlation functions are an
integral part of the theory. They are generated by evanescent operators with anomalous dimension $-4u_*$ at one-loop
order when the fermion flavor number is small. We argue that the negative norm states are a general feature of
fermionic theories in non-integer dimensions. 

It is now clear that unitarity violation occurs in both the scalar and fermionic case. In addition, a recent study also reveals that 
unitarity is violated in non-integer dimensional non-relativistic conformal field theory~\cite{Pal:2018idc}, where unitarity is defined 
by the notion of reflection positivity. Therefore, it seems that unitarity violation is a general property of CFTs in
non-integer dimensions. 

We can't see any way to consistently remove these negative norm states from the fermionic field theory in non-integer
dimensions. They have no effect, however, on theories in integer dimensions where all the negative norm states vanish.

One should mention, however, that although the loss of unitarity prohibits imposing extra constraints while applying the
bootstrap technique, the ``non-unitary bootstrap" technique, which has no reliance on unitarity, still works~\cite{Gliozzi:2013ysa, Gliozzi:2014jsa, Gliozzi:2015qsa}.

It would be a natural extension of our current study to compute the two-loop anomalous dimension matrix and investigate how
the operators in the conformal basis at the two loop order further classify the negative and positive norm states. It would be interesting to investigate the appearance of negative norm states in other fermion/scalar conformal field theories as well.


\vskip5mm

\section*{Acknowledgements}
The authors are grateful for insightful discussions with Vladimir Braun and Alexander Manashov. Y. J. acknowledges the
Deutsche Forschungsgemeinschaft for support under grant BR 2021/7-1.



\appendix


\section{General Formulae}

We provide some key steps in our calculations. The Feynman diagrams in Fig.~\ref{fig:loops} are translated into $T_1^{m,n}$ and
$T_2^{m,n}$ in Eq.~\eqref{eq:DetlaT} as
  \begin{align}
  T_1^{m,n}&=\frac{\delta_{m,n}}{x^4\tr^2(\mathbb{I}_d)(m!)^2}\left[\tr\big(\Gamma^{(m)}_\mu\slashed{x}\,\Gamma^{(m)}_\nu\slashed{x}\big)\right]^2 ,\notag\\
  T_2^{m,n}&=\frac{-1}{x^4\tr(\mathbb{I}_d)m!n!}\tr\big(\slashed{x}\,\Gamma^{(m)}_\mu\slashed{x}\,\Gamma^{(n)}_\nu\slashed{x}\,\Gamma_{(m)}^\mu\slashed{x}\,\Gamma_{(n)}^\nu\big)\,
 \end{align}
 where $\Gamma^{(m)}_{\mu} $ and alike are given in Eq.~(\ref{eq:Gammadefs}).
 Two identities which proved to be the most useful in our study are
 \begin{align}
\label{eq:gG1}
&\gamma^\nu\Gamma^{\mu_1\ldots\mu_n}=\Gamma^{\nu\mu_1\ldots\mu_n}+\sum^{n}_{i=1}(-1)^{i+1}g^{\nu\mu_i}\Gamma^{\mu_1\ldots\widehat\mu_i\ldots\mu_n}\, ,\\
\label{eq:gG2}
&(-1)^n\Gamma^{\mu_1\ldots\mu_n}\gamma^\nu=\Gamma^{\nu\mu_1\ldots\mu_n}+\sum^n_{i=1}(-1)^ig^{\nu\mu_i}\Gamma^{\mu_1\ldots\widehat\mu_i\ldots\mu_n}.
\end{align}
Here $\widehat\mu_i$ denotes that the index $\mu_i$ is omitted. These equations are a consequence of the basic
anti-commutation relation between gamma matrices and the antisymmetric structure of $\Gamma_{(n)}^{\mu}$. By  combining the
latter two equations one finds, 
\be
2\Gamma^{\nu\mu_1\ldots\mu_n}=\gamma^\nu\Gamma^{\mu_1\ldots\mu_n}+(-1)^n\Gamma^{\mu_1\ldots\mu_n}\gamma^\nu\, .
\label{eq:nuGm} 
\ee 
The general formula for contracting the antisymmetrized products of gamma matrices read, 
\be
\Gamma^{\nu_1\ldots\nu_m\mu_n\ldots\mu_1}\Gamma_{\mu_1\ldots\mu_n}=\prod^{n-1}_{i=0}(d-m-i)\Gamma^{\nu_1\ldots\nu_m}\, .
\label{eq:GContract} \ee Finally, we have \be x_a\,x_b\,\Gamma^{\mu_1\ldots a\ldots b\ldots\nu_m}=0\, , \label{eq:xxG} 
\ee 
as a direct consequence of the definition of $\Gamma^ {\mu_1\ldots a\ldots b\ldots\nu_m}$.

We then calculate $T_1^{m,n}$ and $T_2^{m,n}$ separately in the following two sections.


\section{$T_1^{m,n}$}
\label{sec:appT2}


The cyclic property of the trace together with the anti-commutation relation between gamma matrices allow us to first conclude
that $T_1^{m,n}=0$ for $m\neq n$. One thusly writes,
 \begin{align}
  T_1^{m,n}&=\frac{\delta_{m,n}}{x^4\tr^2(\mathbb{I}_d)(m!)^2}\left[\tr\big(\Gamma^{(m)}_\mu\slashed{x}\,\Gamma^{(m)}_\nu\slashed{x}\big)\right]^2 ,\notag\\
  &=B_{m}\,\frac{\delta_{m,n}}{(m!)^2}\, .
 \end{align}
Then we note
\begin{align}
B_m&=\frac{1}{\tr^2(\mathbb{I}_d)}\tr\big(\Gamma^{(m)}_\mu\Gamma^{(m)}_\nu\big)\tr\big(\Gamma_{(m)}^\mu\Gamma_{(m)}^\nu\big)\, ,
\end{align}
which can be proven by the cyclic property of the trace together with Eq.~(\ref{eq:gG1}) and Eq.~(\ref{eq:xxG}).

By setting the default ordering of $\Gamma^{m}_\mu$ and $\Gamma^{(m)}_\nu$ to be $\mu_1,\ldots,\mu_m$ and $\nu_1,\ldots\nu_m$
while noting $\mu_i\neq\mu_j$ and $\nu_i\neq\nu_j$ for $i\neq j$, $B_m$ can then be rewritten as
\begin{align}
\label{eq:R}
B_m&=\frac1{\tr^2(\mathbb{I}_d)}\tr(\gamma_{\mu_1\ldots\mu_m}\gamma_{\nu_1\ldots\nu_m})\tr(\gamma^{\mu_1\ldots\mu_m}\gamma^{\nu_1\ldots\nu_m})\notag\\
&=R_{\mu_1\ldots\nu_m}R^{\mu_1\ldots\nu_m}\, .
\end{align}
By moving $\gamma_{\mu_1}$ to the right of the product in Eq.~\eqref{eq:R} and using the cyclic property of the trace one
finds
\begin{align}
R_{\mu_1\ldots\nu_m}
&=\frac{(-1)^{m-1}}{\tr(\mathbb{I}_d)}\sum^m_{i=1}(-1)^{i+1}g_{\mu_1\nu_i}\notag\\
&\qqqquad\qqqquad\,\,\,\times\tr(\gamma_{\mu_2\ldots\mu_m}\gamma_{\nu_1\ldots\widehat\nu_i\ldots\nu_m}),
\label{eq:ggmm}
\end{align}
where again $\widehat\nu_i$ denotes the omitted index.

Here we have reduced a trace with $2m$ indices to a sum of traces with $2(m-1)$ indices. By repeatedly applying
Eq.~(\ref{eq:ggmm}) one can reduce the trace with $2m$ indices to $\tr(\mathbb{I}_d)$ with an appropriate combination of
coefficients. It is clear that each further reduction step produces one extra summation and one extra set of
$g^{\mu_{i_k}\nu_{j_l}}$ with appropriate sign in front. To this end, we write,
\begin{align}
R_{\mu_1\ldots\nu_m}&=(-1)^{\frac m2(m-1)}\tr(\mathbb{I}_d)\bigg\{\sum^m_{i_1=1} \ldots\sum^m_{i_m=1}g_{\mu_1\nu_{i_1}}\ldots\notag\\
&\qqqquad\qquad\times g_{\mu_m\nu_{i_m}}\Omega(i_1,\ldots, i_m)\bigg\},
\end{align}
where the overall factor $(-1)^{\frac m2(m-1)}$ is accumulated from repeated use of Eq.~(\ref{eq:ggmm}) and
$\Omega(i_1,\ldots,i_m)\in\{0,\pm1\}$. Since each index $\nu_k$ appears only once in the trace, one straightforwardly
concludes that $\Omega(i_1,\ldots,i_k,\ldots,i_k,\ldots,i_m)=0$. A more detailed analysis also reveals that $\Omega(i_1,\ldots,
i_k,i_{k+1},\ldots, i_m)=-\Omega(i_1,\ldots, i_{k+1},i_{k},\ldots, i_m)$ which is a property inherited from the antisymmetric
nature of $\Gamma^{(m)}$. Finally by noting that $\Omega(1,\ldots, m)=1$, which corresponds to eliminate $\gamma_{\nu_1}\,,
\gamma_{\nu_2}, \ldots, \gamma_{\nu_m}$ in order, one identifies,
$$
\Omega(i_1,\ldots,i_m)=\epsilon_{i_1\ldots i_m}\, .
$$
Therefore one finds,
 \begin{align}
 B_{m}=\sum^m_{\substack{i_1,\ldots,i_m=1\\j_1,\ldots,j_m=1}} g^{\mu_{i_1}}_{\mu_{j_1}}\ldots g^{\mu_{i_m}}_{\mu_{j_m}}\epsilon_{i_1\ldots i_m}\epsilon^{j_1\ldots j_m},
 \end{align}
 with
 \begin{align}
 \sum^m_{\substack{i_1,\ldots,i_m=1\\j_1,\ldots,j_m=1}}\equiv\sum^m_{i_1=1}\ldots\sum^m_{i_m=1}\sum^m_{j_1=1}\ldots\sum^m_{j_m=1}\, .
 \end{align}
This summation can be worked out by dividing the general case into two scenarios:
\begin{align}
i_k&=j_k=m\, ,\quad k=1,\ldots m\, ,\notag\\
\text{or}\quad
i_k&=j_l=m\, ,k\neq l\, ,\quad m,l=1,\ldots m\, .
\end{align}
The summation is then easily carried out and leads to a recurrence relation for $B_m$,
 \begin{align}
 \label{eq:T1Arec}
 B_{m}=m(d-m+1)B_{m-1}\, ,
 \end{align}
 for $m > 1$.
Combined with the initial condition of the sequence $B_0=1$, we obtain the final expression for $T_1^{m,n}$,
\begin{align}
T_1^{m,n}&=\frac{\Gamma(d+1)}{m!\,\Gamma(d-m+1)}\,\delta_{m,n}\, .
\end{align}

It is clear that $T_1^{m,n}$ vanishes for theories in even $d$ dimensions if $m > d$. This observation
is in accordance with the fact that there are $d$ numbers of gamma matrices in even $d$ dimensions and
consequently, the antisymmetrized product of $m$ gamma matrices $\Gamma^{(m)}_\mu$ vanishes if $m > d$. 
In odd dimension $d$, $\Gamma^{(d)}$ is no longer an independent matrix since $\Gamma^{(d)}\propto\Gamma^{(0)}$ and therefore this redundancy must be removed ``by hand".
 If the dimension $d$ is no longer an integer, however, then $T_1^{m,n}$ never vanishes and can take negative values.


\section{$T_2^{m,n}$}


We proceed to calculate $T_2^{m,n}$ next:
\begin{align}
T_2^{m,n}&=\frac{-1}{x^4\tr(\mathbb{I}_d)m!n!}\tr\big(\slashed{x}\,\Gamma^{(m)}_\mu\slashed{x}\,\Gamma^{(n)}_\nu\slashed{x}\,\Gamma_{(m)}^\mu\slashed{x}\,\Gamma_{(n)}^\nu\big)\,
\end{align}
which can be simplified using Eq.~(\ref{eq:gG1})--(\ref{eq:xxG}) as,
\begin{align}
T_2^{m,n}&=\frac{-1}{\tr(\mathbb{I}_d)m!n!}\tr\left(\Gamma^{(m)}_\mu\Gamma^{(n)}_\nu\Gamma_{(m)}^\mu\Gamma_{(n)}^\nu\right)\, .
\end{align}
Then again with Eq.~(\ref{eq:nuGm}) and Eq.~(\ref{eq:GContract}) one finds,
\begin{align}
T_2^{m,n}&=\frac{-1}{2\tr(\mathbb{I}_d)m!n!}
\tr\Big[\Gamma^{\nu_2\ldots\nu_n}\Gamma^{(m)}\gamma_{(n)}\Gamma_{(m)}\gamma^{\nu_1}\notag\\
&\quad+(-1)^{m+n-1}(d-2m)\Gamma^{(n-1)}\Gamma^{(m)}\gamma_{(n-1)}\Gamma_{(m)}\Big]\notag\\
&=\frac{-1}{2\tr(\mathbb{I}_d)m!n!}(s_1+s_2)\,,
\label{eq:G4s12}
\end{align}
where $\gamma_{(n-k)}=\gamma_{\nu_1\ldots\nu_{n-k}}=\gamma_{\nu_1}\ldots\gamma_{\nu_{n-k}}$ is a product of gamma matrices of
standard ordering. Then,
\begin{align}
s_1
&=\tr\Big\{2\sum^{n-1}_{i=1}(d-n+2)\Gamma^{(n-2)}\Gamma^{(m)}\gamma_{(n-2)}\Gamma_{(m)}\notag\\
&\quad+(-1)^{m+n-1}(d-2m)\Gamma^{(n-1)}\Gamma^{(m)}\gamma_{(n-1)}\Gamma_{(m)}\Big\}\, .
\end{align}
One then obtains a recursion relation for $T_2^{m,n}$,
\begin{align}
\label{eq:T2rec}
T_2^{m,n}&=(-1)^{m+n-1}(d-2m)T_2^{m,n-1}\notag\\
&\quad+(n-1)(d-n+2)T_2^{m,n-2}\, ,
\end{align}
with boundary conditions,
\begin{align}
T_2^{m,0}&=-\frac{(-1)^{\frac m2(m-1)}}{m!}\,\frac{\Gamma(d+1)}{\Gamma(d-m+1)}\, ,\notag\\
T_2^{m,1}&=-\frac{(-1)^{\frac m2(m+1)}}{m!}\,\frac{(d-2m)\,\Gamma(d+1)}{\Gamma(d-m+1)}\, .
\label{eq:bc}
\end{align}

The recurrence relation can be expressed as
\begin{align}
\label{eq:T2amn}
T_2^{m,n}&=-\frac12\,i^{m(m+1)+n(n+1)}\,a_{m,n}\, ,
\end{align}
where $a_{m,n}$ are the coefficients of the generating function
\begin{align}
\label{eq:generating}
F(x,y)&=\sum_{m,n=0}^\infty a_{m,n}\,x^m\,y^n\, .
\end{align}

Then the recursive behavior in Eq.~(\ref{eq:T2rec}) is inherited by $a_{m,n}$ and combined with the boundary conditions
Eq.~(\ref{eq:bc}), the generating function is found to be
\begin{align}
F(x,y)&=(1-x+y+xy)^d+(1+x-y+xy)^d\notag\\
&-(1+x+y-xy)^d+(1-x-y-xy)^d\, .
\end{align}


\section{Anomalous Dimensions}

\label{sec:AD}


The Feynman diagrams in the first row of Fig.~\ref{fig:AMD} has a divergent part which reads,
\begin{align}
I_1^{(m)}&=(g_1^*)^2\frac{m-2}{16\pi^2\epsilon}\,(-1)^m{\cal O}^{(m)}\, .
\end{align}
The Feynman diagrams in the last row of Fig.~\ref{fig:AMD} together yield,
\begin{align}
I_2^{(m)}&\stackrel{\rm div}{=}\frac{(g_1^*)^2}{64\pi^2m!\epsilon}\bar\Psi(x)[\Gamma^{(m)},\gamma_\mu]\Psi(x)\notag\\
&\qqqquad\qquad\times\bar\Psi(x)[\Gamma_{(m)},\gamma^\mu]\Psi(x)\, ,
\end{align}
which requires some algebra. For odd $m$, we get,
\begin{align}
I_2^{(m={\rm odd})}&=\frac{(g_1^*)^2(m+1)}{16\pi^2\epsilon}{\cal O}^{(m+1)}\, ,
\end{align}
while for the even case, we have,
\begin{align}
I_2^{(m={\rm even})}&=\frac{(g_1^*)^2(5-m)}{16\pi^2\epsilon}{\cal O}^{(m-1)}\, .
\end{align}
The Feynman diagrams in Fig.~\ref{fig:AMDphy} yield a divergent piece which reads,
\begin{align}
I_3^{(m)}&=\frac{(g_1^*)^2 {\cal O}^{(0)}}{16\pi^2\epsilon}\left[\frac{(-1)^{m(m-1)/2}}{m!}\prod^{m-1}_{i=0}(4-i)-{N_f}\delta_{m,0}\right].
\end{align}
Consequently, the operator renormalization matrix is found to be
\begin{align}
Z_{\cal O}^{m,n}&=I_1+I_2+I_3\notag\\
&=\frac{(g_1^*)^2}{16\pi^2\epsilon}\Big[(n-2)(-1)^n\delta_{m,n}+n\delta_{m,n-1}{\rm mod}(m,2)\notag\\
&+(4-n)\delta^{m,n+1}{\rm mod}(n,2)-{N_f}\delta_{m,0}\delta_{n,0}\notag\\
&+\frac{4!(-1)^{m(m-1)/2}}{m!(4-m)!}\delta_{n,0}\Big]\, .
\end{align}
The one-loop renormalization of the fermion self energy in GNY model reads
\begin{align}
Z_\Psi^{m,n}&=1+\frac{(g_1^*)^2}{32\pi^2\epsilon}\delta_{m,n}
\end{align}
and therefore one obtains the one-loop anomalous dimensions matrix
\begin{align}
&{\boldsymbol\gamma_{{\cal O}}^{m,n}}=\frac{d \alpha_1}{d\ln\mu}\frac{\partial}{\partial \alpha_1}\ln(Z_{{\cal{O}}}Z^{-2}_{\Psi})\notag\\
&\quad=2u_*\bigg[(1-(-1)^n(n-2))\delta_{m,n}-n{\delta_{m,n-1}}{\rm mod}(m,2)\notag\\
&\qqquad+(n-4)\delta_{m,n+1}{\rm mod}(n,2)+{N_f}\delta_{m,0}\delta_{n,0}\notag\\
&\qqquad-\frac{4!(-1)^{m(m-1)/2}}{m!(4-m)!}\delta_{n,0}\bigg]\, .
\end{align}



\begin{thebibliography}{99}




\bibitem{Hogervorst:2015akt}
  M.~Hogervorst, S.~Rychkov and B.~C.~van Rees,
  Phys.\ Rev.\ D {\bf 93} (2016) no.12,  125025
  [arXiv:1512.00013 [hep-th]].




\bibitem{Polyakov:1974gs}
  A.~M.~Polyakov,
  Zh.\ Eksp.\ Teor.\ Fiz.\  {\bf 66} (1974) 23
   [Sov.\ Phys.\ JETP {\bf 39} (1974) 9].
  



\bibitem{Ferrara:1973yt}
  S.~Ferrara, A.~F.~Grillo and R.~Gatto,
  Annals Phys.\  {\bf 76} (1973) 161.




\bibitem{Rattazzi:2008pe}
  R.~Rattazzi, V.~S.~Rychkov, E.~Tonni and A.~Vichi,
  JHEP {\bf 0812} (2008) 031
  [arXiv:0807.0004 [hep-th]].




\bibitem{Rychkov:2011et}
  S.~Rychkov,
 ``Conformal Bootstrap in Three Dimensions?,''
  arXiv:1111.2115 [hep-th].




\bibitem{ElShowk:2012ht}
  S.~El-Showk, M.~F.~Paulos, D.~Poland, S.~Rychkov, D.~Simmons-Duffin and A.~Vichi,
  Phys.\ Rev.\ D {\bf 86} (2012) 025022
  [arXiv:1203.6064 [hep-th]].
  
  


\bibitem{Kos:2013tga}
  F.~Kos, D.~Poland and D.~Simmons-Duffin,
  JHEP {\bf 1406} (2014) 091
  [arXiv:1307.6856 [hep-th]].
  



\bibitem{Rychkov:2009ij}
  V.~S.~Rychkov and A.~Vichi,
  Phys.\ Rev.\ D {\bf 80} (2009) 045006
  [arXiv:0905.2211 [hep-th]].




\bibitem{Caracciolo:2009bx}
 F.~Caracciolo and V.~S.~Rychkov,
  Phys.\ Rev.\ D {\bf 81} (2010) 085037
  [arXiv:0912.2726 [hep-th]].




\bibitem{Poland:2010wg}
  D.~Poland and D.~Simmons-Duffin,
  JHEP {\bf 1105} (2011) 017
  [arXiv:1009.2087 [hep-th]].




\bibitem{Rattazzi:2010gj}
  R.~Rattazzi, S.~Rychkov and A.~Vichi,
  Phys.\ Rev.\ D {\bf 83} (2011) 046011
  [arXiv:1009.2725 [hep-th]].




\bibitem{Rattazzi:2010yc}
  R.~Rattazzi, S.~Rychkov and A.~Vichi,
  J.\ Phys.\ A {\bf 44} (2011) 035402
  [arXiv:1009.5985 [hep-th]].




\bibitem{Vichi:2011ux}
  A.~Vichi,
  JHEP {\bf 1201} (2012) 162
  [arXiv:1106.4037 [hep-th]].
  



\bibitem{Nakayama:2014yia}
  Y.~Nakayama and T.~Ohtsuki,
  Phys.\ Lett.\ B {\bf 734} (2014) 193
  [arXiv:1404.5201 [hep-th]].




\bibitem{Li:2016wdp}
  Z.~Li and N.~Su,
  JHEP {\bf 1704} (2017) 098
  [arXiv:1607.07077 [hep-th]].





\bibitem{Bae:2014hia}
  J.~B.~Bae and S.~J.~Rey,
  arXiv:1412.6549 [hep-th].



\bibitem{Poland:2011ey}
  D.~Poland, D.~Simmons-Duffin and A.~Vichi,
  JHEP {\bf 1205} (2012) 110
  [arXiv:1109.5176 [hep-th]].




\bibitem{ElShowk:2012hu}
  S.~El-Showk and M.~F.~Paulos,
  Phys.\ Rev.\ Lett.\  {\bf 111} (2013) no.24,  241601
  [arXiv:1211.2810 [hep-th]].



\bibitem{Wilson:1971dc}
  K.~G.~Wilson and M.~E.~Fisher,
  Phys.\ Rev.\ Lett.\  {\bf 28} (1972) 240.




\bibitem{Wilson:1972cf}
  K.~G.~Wilson,
  Phys.\ Rev.\ D {\bf 7} (1973) 2911.

 


\bibitem{Kompaniets:2017yct}
  M.~V.~Kompaniets and E.~Panzer,
  Phys.\ Rev.\ D {\bf 96}, no. 3, 036016 (2017)
  [arXiv:1705.06483 [hep-th]].
  



\bibitem{El-Showk:2013nia}
  S.~El-Showk, M.~Paulos, D.~Poland, S.~Rychkov, D.~Simmons-Duffin and A.~Vichi,
  Phys.\ Rev.\ Lett.\  {\bf 112} (2014) 141601
  [arXiv:1309.5089 [hep-th]].




\bibitem{Liendo:2012hy}
  P.~Liendo, L.~Rastelli and B.~C.~van Rees,
  JHEP {\bf 1307} (2013) 113
  [arXiv:1210.4258 [hep-th]].




\bibitem{Hogervorst:2014rta}
  M.~Hogervorst, S.~Rychkov and B.~C.~van Rees,
  Phys.\ Rev.\ D {\bf 91} (2015) 025005
  [arXiv:1409.1581 [hep-th]].




\bibitem{Dugan:1990df}
  M.~J.~Dugan and B.~Grinstein,
  Phys.\ Lett.\ B {\bf 256} (1991) 239.




\bibitem{ZinnJustin:1991yn}
  J.~Zinn-Justin,
  Nucl.\ Phys.\ B {\bf 367} (1991) 105.




\bibitem{Hasenfratz:1991it}
  A.~Hasenfratz, P.~Hasenfratz, K.~Jansen, J.~Kuti and Y.~Shen,
  Nucl.\ Phys.\ B {\bf 365} (1991) 79.




\bibitem{Moshe:2003xn}
  M.~Moshe and J.~Zinn-Justin,
  Phys.\ Rept.\  {\bf 385} (2003) 69
  [hep-th/0306133].
  
  


\bibitem{Zerf:2017zqi}
  N.~Zerf, L.~N.~Mihaila, P.~Marquard, I.~F.~Herbut and M.~M.~Scherer,
  Phys.\ Rev.\ D {\bf 96}, no. 9, 096010 (2017)
  [arXiv:1709.05057 [hep-th]].




\bibitem{Herrlich:1994kh}
  S.~Herrlich and U.~Nierste,
  Nucl.\ Phys.\ B {\bf 455}, 39 (1995)
  [hep-ph/9412375].




\bibitem{Collins:1984xc}
  J.~C.~Collins,
  \emph{Renormalization}. Cambridge University Press, 1984.




\bibitem{Vasiliev:1995qj}
  A.~N.~Vasiliev, S.~E.~Derkachov and N.~A.~Kivel,
  Theor.\ Math.\ Phys.\  {\bf 103} (1995) 487
   [Teor.\ Mat.\ Fiz.\  {\bf 103} (1995) 179].
 



\bibitem{Fei:2016sgs}
  L.~Fei, S.~Giombi, I.~R.~Klebanov and G.~Tarnopolsky,
  PTEP {\bf 2016} (2016) no.12,  12C105
  [arXiv:1607.05316 [hep-th]].




\bibitem{Gliozzi:2013ysa}
  F.~Gliozzi,
  Phys.\ Rev.\ Lett.\  {\bf 111} (2013) 161602
  [arXiv:1307.3111 [hep-th]].




\bibitem{Gliozzi:2014jsa}
  F.~Gliozzi and A.~Rago,
  JHEP {\bf 1410} (2014) 042
  [arXiv:1403.6003 [hep-th]].


\bibitem{Gliozzi:2015qsa}
  F.~Gliozzi, P.~Liendo, M.~Meineri and A.~Rago,
  JHEP {\bf 1505} (2015) 036
  [arXiv:1502.07217 [hep-th]].



\bibitem{Pal:2018idc}
  S.~Pal,
  arXiv:1802.02262 [hep-th].




\end{thebibliography}
\end{document}